\documentclass[prb, twocolumn,showpacs,preprintnumbers,amsmath,amssymb, floatfix]{revtex4}

\usepackage{graphicx,amsmath,amssymb}
\usepackage[usenames]{color}

\begin{document}
\newcommand{\kk}{{\bf k}}
\newcommand{\q}{{\bf q}}
\newcommand{\HH}{\mathcal{H}}

\title{Mean-Field Analysis of Intra-Unit-Cell Order in the Emery Model of the CuO$_2$ Plane}

\author{Mark H. Fischer}
\author{Eun-Ah Kim}
\affiliation{%
Department of Physics, Cornell University, Ithaca, New York 14853, USA
}%

\date{\today}

\begin{abstract}
  Motivated by recent experiments on high-$T_c$ cuprate
  superconductors pointing towards
  intra-unit-cell order in the
  pseudogap phase, 
  we investigate three distinct intra-unit-cell-ordering possibilities: 
nematic, nematic-spin-nematic, and current-loop order.
The first two are Fermi-surface instabilities involving 
 a spontaneous charge and magnetization imbalance between the two oxygen sites in
 the unit cell, respectively, while the third describes circulating currents within the unit cell.
  We analyze the three-band Emery model of a
  single CuO$_2$ layer including various on-site and nearest-neighbor
  interactions within a self-consistent mean-field approach.
 We show how these on-site and further-neighbor
  repulsions suppress or enhance particular IUC orders.
  In particular, we show that the attractive interactions
  necessary for nematic and nematic-spin-nematic
  orders in one-band models have their natural
microscopic origin in the O-O on-site and nearest-neighbor repulsions
in the three-band model.
Finally, we find that while the nematic and nematic-spin-nematic orders 
cannot coexist in this framework, 
the loop-current order can coexist with nematic order.
\end{abstract}

\maketitle
\section{Introduction}
\label{sec:intro}
Experimental evidence for various types of symmetry breaking in the pseudogap
region of the phase diagram of the high-$T_c$ cuprate superconductors
has been accumulating in recent years. 
Neutron scattering experiments discovered a subtle
staggered magnetic order in the pseudo-gap region of
YBCO\cite{fauque:2006} and Hg-compounds\cite{li:2008}
that could be accounted for by either so-called nematic-spin-nematic
order\cite{oganesyan:2001, wu:2007} or circulating current
loops\cite{varma:2006b}. On the other hand, neutron
scattering~\cite{hinkov:2008} and Nernst effect~\cite{daou:2010}
measurements on YBCO as well as SI-STM on BSCCO~\cite{lawler:2010}
point towards an electronic nematic state. All these states retain the
translational symmetry of the underlying crystal and can thus
naturally be described by breaking intra-unit-cell (IUC) symmetries. Hence,
identifying mechanisms for these symmetry-breaking possibilities and
understanding their competition is crucial for understanding the
nature of the pseudogap phase.  

Theoretical investigations of translationally-invariant IUC order 
have so far been focussing on one particular ordering at a time within simplified
models each aimed at the ordering of interest.
Nematic and nematic-spin-nematic order have only been studied in one-band models\cite{yamase:2000c,oganesyan:2001, wu:2007, kee:2003,metzner:2003, yamase:2005,
halboth:2000, gull:2009,okamoto:2010} or in the extreme limit of infinite interactions\cite{kivelson:2004}. Loop currents, being more dependent on an IUC picture, have been studied in a mean-field picture with additional assumptions\cite{varma:2006b} or numerically on small clusters or ladders\cite{chudzinski:2007, greiter:2008, weber:2009}.

\begin{figure}[b]
  \begin{center}
    \includegraphics{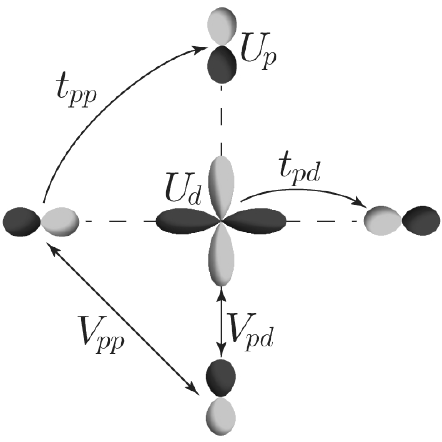}
  \end{center}
  \caption{The unit cell of the CuO$_2$ plane with the copper $d_{x^2-y^2}$ in the middle surrounded by the oxygen $p_x$ and $p_y$ orbitals. Also shown are the different hopping as well as interaction parameters used in the Emery model.}
  \label{fig:unitcell}
\end{figure}

Here, we aim at a comprehensive investigation of IUC-ordering
possibilities suggested by recent experiments.~\footnote{Other IUC-ordering possibilities were considered e.g. by Sun et al.~\cite{sun:2008}}
A theoretical description of intra-unit-cell order
should be based on a microscopic model
that allows for structures within the unit-cell.
We thus start from a three-band model for the CuO$_2$ plane, the so-called Emery
model\cite{emery:1987}, and consider various on-site and nearest-neighbor (nn) interactions
(see Fig.~\ref{fig:unitcell}).
We analyze three
distinct IUC orders: nematic, nematic-spin-nematic,
 and loop currents.
 These phases can be distinguished by the respective symmetries they break, both of the point group $D_{4h}$ and time reversal, as is summarized in Tab.~\ref{tab:sym}. 
For simplicity, only fourfold rotations, inversion, time reversal and combinations of these are shown. Within a self-consistent mean-field scheme we analyze and compare the origins of these phases and compare the influence
 of the different model parameters on them.
In addition, we show how the O-O on-site and nn interactions result in
effective interactions of $d_{x^2-y^2}$ symmetry in one-band models, thus naturally leading to nematic or nematic-spin-nematic order.
\begin{table}[b!]
  \centering
  \begin{tabular}{c|c|c|c|c|c}
    & $C_4$ & $\mathcal{I}$ & $\mathcal{T}$ & $C_4\circ \mathcal{T}$ & $\mathcal{I}\circ\mathcal{T}$\\
    \hline
    \hline
    nematic &$\quad\times\quad$ & $\quad\phantom{-}\quad$ & $\quad\phantom{-}\quad$ & $\quad\times\quad$ &$\quad\phantom{-}\quad$\\
    \hline
    nematic-spin-nematic &$\times$  &  & $\times$ & $ $ & $ $\\
    \hline
    $\Theta_{II}$ loop current & $\times$ & $\times$ & $\times$ & $\times$ &
  \end{tabular}
  \caption{The broken symmetries distinguishing the different IUC orderings with $\times$ denoting symmetries broken in the respective phase. For simplicity, we restrict the table to the fourfold rotation $C_4$, the inversion $\mathcal{I}$, time-reversal operation $\mathcal{T}$ as well as combinations thereof.}
  \label{tab:sym}
\end{table}

This paper is organized as follows: After introducing the Emery model describing the CuO$_{2}$ plane, section \ref{sec:static} deals with nematic and nematic-spin-nematic IUC order through a decoupling of various interactions in the Hartree channel within self-consistent mean-field theory.  
In addition, we compare the three-band with the one-band model by focussing on the (partially filled) lowest of the three bands. 
Section~\ref{sec:loop} examines IUC loop currents by decoupling the nearest-neighbor interactions of the full Emery model in the Fock channel. Finally, section~\ref{sec:discussion} compares the results for the different orderings and concludes.  
\section{Model}
\label{sec:model}
The kinetic part of the Emery model~\cite{emery:1987} describing hopping
of holes in the CuO$_2$ plane is
\begin{multline}
  \HH_{0} = -t_{pd}\sum_{i,s}\sum_\nu(\hat{d}^{\dag}_{i,s}\hat{p}^{\phantom{\dag}}_{ i+\hat{\nu}/2, s} + {\rm h.c.})\\
  - t_{pp}\sum_{i, s}\sum_{\langle\nu,\nu'\rangle}(\hat{p}^{\dag}_{ i+\hat{\nu}/2, s}\hat{p}^{\phantom{\dag}}_{ i+\hat{\nu}'/2, s} + {\rm h.c.})\\ 
  - \mu \sum_{i,s}\hat{n}^{d}_{i,s} - \frac12(\mu - \Delta)\sum_{i,s}\sum_{\nu}\hat{n}^{p}_{i+\hat{\nu}/2,s}
  \label{eq:hopping}
\end{multline}
with $t_{pd}$ and $t_{pp}$ the Cu-O and O-O hopping integrals. Here, $\hat{d}^{\dag}_{i,s}$ creates a hole
in the copper $d_{x^2-y^2}$ orbital at site $i$ with spin $s$,
$\hat{p}^{\dag}_{ i+\hat{\nu}/2, s}$ creates a hole in the oxygen $p_\nu$ orbital
at the site $i+\hat{\nu}/2$
for $\nu = x, y$, and
$\hat{n}^{d}_{i,s}$,
$\hat{n}^{p}_{ i+\hat{\nu}, s}$
are the corresponding number operators.
The Cu sites $i$ form a square
lattice with unit vectors $\hat{x}$ and $\hat{y}$, and the total number of lattice sites is $N$. The chemical potential $\mu$
and the charge transfer energy $\Delta$ control the total and relative
Cu/O hole densities, and $\langle\nu, \nu'\rangle$ point to neighboring oxygen sites.

In addition, we consider the interaction Hamiltonian including on-site interactions with strengths $U_{d}$ and $U_{p}$ as well as nn interactions, $V_{pd}$ and $V_{pp}$,
\begin{multline}
  \HH' = U_{d}\sum_{i}\hat{n}_{i\uparrow}^{d}\hat{n}_{i\downarrow}^{d} + \frac{U_{p}}{2}\sum_{i, \nu}\hat{n}_{ i+\hat{\nu}/2,\uparrow}^{p}\hat{n}_{ i+\hat{\nu}/2,\downarrow}^{p}\\
  + V_{pd}\sum_{i, \nu}\sum_{s, s'}\hat{n}_{i,s}^{d}\hat{n}_{ i+\hat{\nu}/2, s'}^{p}\\
  + V_{pp}\sum_{i}\sum_{\langle\nu,\nu'\rangle}\sum_{s,s'}\hat{n}_{ i+\hat{\nu}/2, s}^{p}\hat{n}_{i+\hat{\nu}/2',s'}^{p}.
  \label{eq:interaction}
\end{multline}
The different orbitals and parameters of the model are shown in
Fig.~\ref{fig:unitcell}. Setting $t_{pd}=1$, we fix the energy scale
in the following.

\section{Nematic and Nematic-Spin-Nematic Order}
\label{sec:static}
For the above introduced Emery model, only the strong coupling limit, taking all interactions to infinity, has been analyzed for nematicity.
Most theoretical investigations of nematic and nematic-spin-nematic order start from a single-band model, where in the weak-coupling limit a quadrupolar\cite{oganesyan:2001, wu:2007, kee:2003} or a forward-scattering interaction\cite{metzner:2003, yamase:2005} is introduced. For systems with a sufficiently high density of states at the Fermi energy, e.g. due to a van Hove singularity, this can lead to a Pomeranchuck instability in the $d$-wave channel. Other studies of the Hubbard model without any additional (long-range) interactions found a nematic instability within a (weak-coupling) RG approach~\cite{halboth:2000}, while DMFT calculations showed that the model maintains $C_{4}$ symmetry,~\cite{gull:2009, okamoto:2010} but becomes very susceptible to weak nematic driving fields (such as lattice distortions) close to the Mott transition.\cite{okamoto:2010}

In the three-band model,
the oxygen-oxygen nn interaction prefers an imbalance in the hole
densities of the neighboring oxygen sites, whereas the oxygen on-site
interaction prefers to spin polarize the oxygen sites. The former interaction
can thus lead to nematic order, the breaking of $C_4$ symmetry, and
the latter to either an overall magnetization on the oxygen sites or a
nematic-spin-nematic order, which is invariant under a combination of
a $C_4$ rotation and time reversal (see Tab.~\ref{tab:sym}). Solving self-consistently the
mean-field equations, we analyze the nematic and nematic-spin-nematic ordering in the following.
\subsection{Mean-Field Theory}
\label{subsec:model}
\begin{figure}[tb]
  \begin{center}
    \includegraphics{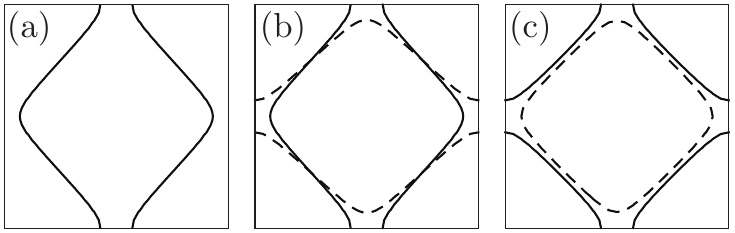}
  \end{center}
  \caption{Static Fermi surface instabilities analyzed in this work:
    (a) Nematic phase breaking $C_{4}$ symmetry, (b)
    nematic-spin-nematic and (c) ferromagnetic instability. 
    In (b) and (c), the solid and dashed lines denote the up- and down-spin band.
    }
  \label{fig:fs}
\end{figure}

In this section, we focus only on symmetry breaking associated
with the hole densities on the oxygen sites $n^{p}_{\nu s} = \langle
\hat{n}^{p}_{i+\hat{\nu}/2 , s}\rangle$.
In the absence of an overall magnetization on the oxygen sites, i.e.
\begin{equation}
  m\equiv (n_{x\uparrow}^p-n_{x\downarrow}^{p}) + (n_{y\uparrow}^p - n_{y\downarrow}^p)=0,
  \label{eq:mag}
\end{equation}
there are  two
distinct ways to break the lattice symmetry within each unit-cell, a nematic order and a
nematic-spin nematic order.
IUC nematic order in this model can be measured in terms of
a spontaneous imbalance between the density of holes at the oxygen sites
\begin{equation}
  \eta\equiv (n_{x\uparrow}^p+n_{x\downarrow}^{p}) -(n_{y\uparrow}^p + n_{y\downarrow}^p),
  \label{eq:eta}
\end{equation}
while nematic-spin-nematic order corresponds to having equal, but opposite magnetization on the two oxygen sites, 
\begin{equation}
  \eta_s\equiv (n_{x\uparrow}^p-n_{x\downarrow}^{p}) -(n_{y\uparrow}^p - n_{y\downarrow}^p).
  \label{eq:etas}
\end{equation}
Fig.~\ref{fig:fs} shows the distorted Fermi surfaces associated with
these ordering possibilities as well as for ferromagnetic ordering for contrast.

We arrive in the following at the mean-field Hamiltonian for nematic or
nematic-spin-nematic order by a standard decoupling of all the
interaction terms in Eq.~\eqref{eq:interaction} in the Hartree
channel. We seek self-consistent solution with $\eta\neq0$ or
$\eta_s\neq0$, respectively, for nematic and nematic-spin-nematic
order. 

\subsubsection{Nematic order}
The mean-field Hamiltonian can be written in momentum space as
\begin{equation}
  \HH_{\rm MF} = \sum_{\kk, s} \hat{C}^{\dag}_{\kk s}\HH_{\kk s}\hat{C}^{\phantom{\dag}}_{\kk s} + f(n^{p}, \eta)
  \label{eq:mfhamiltonian}
\end{equation}
with $\hat{C}^{\dag}_{\kk s} = (\hat{p}^{\dag}_{x\kk s}, \hat{p}^{\dag}_{y\kk s}, \hat{d}^{\dag}_{\kk s})$,
\begin{equation}
  \HH_{\kk s} = \left(\begin{array}{ccc} \xi_{x}& \gamma_2(\kk)& \gamma_1(k_x)\\ \gamma_2(\kk)&\xi_{y}& \gamma_1(k_y)\\ \gamma_1(k_x)& \gamma_1(k_y)&\xi_{d}\end{array}\right)
  \label{eq:h0k}
\end{equation}
and
\begin{equation}
  \frac{f(n^{p}, \eta)}{N} = -\tilde{U}_{p} \frac{(n^{p})^2}{8} + \tilde{V}_{pp}\frac{\eta^{2}}{8} - \tilde{U}_{d}\frac{(n-n^{p})^2}{4}.
  \label{eq:mutilde}
\end{equation}
Here, $n$ is the total
density of holes, $n^p$ is the total
density of holes on the oxygen sites, i.e., 
\begin{equation}
  n^p\equiv (n_{x\uparrow}^p+n_{x\downarrow}^{p}) + (n_{y\uparrow}^p + n_{y\downarrow}^p),
  \label{eq:np}
\end{equation}
and the nematic order parameter $\eta$ is defined in Eq.\eqref{eq:eta}.
The elements of the matrix~\eqref{eq:h0k} are given by
\begin{eqnarray}
  \gamma_1(k_i) &=& -2t_{pd}\cos\frac{k_i}{2}, \label{eq:gammas}\\
  \gamma_2(\kk) &=& -4t_{pp}\cos\frac{k_x}{2}\cos\frac{k_y}{2},
\end{eqnarray}
and
\begin{eqnarray}
  \xi_{x} &=& \Delta + \tilde{U}_{p} \frac{n^{p}}{4} - \tilde{V}_{pp}\frac{\eta}{4} - \mu,\\
  \xi_{y} &=& \Delta + \tilde{U}_{p} \frac{n^{p}}{4} + \tilde{V}_{pp}\frac{\eta}{4} - \mu,\\
  \xi_{d} &=& \tilde{U}_{d} \frac{(n-n^p)}{2} - \mu,
  \label{eqn:diag}
\end{eqnarray}
and also, we introduced the effective interaction parameters
\begin{eqnarray}
  \tilde{U}_{p} &=& U_{p} + 8V_{pp} - 8 V_{pd},\label{eq:Uptilde}\\
  \tilde{V}_{pp} &=&  8V_{pp} - U_{p},\label{eq:Vptilde}\\
  \tilde{U}_{d} &=& U_{d} - 4V_{pd}.
  \label{eq:Udtilde}
\end{eqnarray}
In addition, we have put all the constant terms, i.e. $2V_{pd}n - V_{pd}n^2$, into the chemical potential $\mu$.
The mean-field Hamiltonian~\eqref{eq:mfhamiltonian} can be diagonalized to yield three bands
each with mixed orbital character and dispersion $\xi_{\alpha\kk s}$, where $\alpha=1,2,3$ is the
band index for the lowest lying and the two upper bands.

In order to self-consistently determine the above introduced mean fields, we look at the grand potential per lattice site
\begin{equation}
  \omega = -\frac{T}{N}\sum_{\alpha, \kk, s} \log[1 + \exp(-\xi_{\alpha \kk s}/T)] + \frac{f(n^{p}, \eta)}N.
  \label{eq:grand}
\end{equation}
For given values of $n^{p}$ and $\eta$, the chemical potential $\mu$ is implicitly given by solving
\begin{equation}
  n = -\frac{\partial\omega}{\partial\mu}=\frac{1}{N}\sum_{\alpha, \kk, s} n_{\rm F}(\xi_{\alpha\kk s})
  \label{eq:implicitmu}
\end{equation}
with the Fermi distribution function $n_{\rm F}(x)=1/(\exp(x/T) +1)$.
Self-consistency equations for $n^p$ and $\eta$ are found by
extremizing the grand potential~\eqref{eq:grand} to be
\begin{equation}
  n^{p} = \frac{4}{N(\tilde{U}_{p} + 2\tilde{U}_{d})}\sum_{\alpha, \kk, s}n_{\rm F}(\xi_{\alpha\kk s})\frac{\partial \xi_{\alpha\kk s}}{\partial n^{p}} + \frac{2\tilde{U}_{d}n}{\tilde{U}_{p} + 2\tilde{U}_{d}}
  \label{eq:selfconn}
\end{equation}
and
\begin{equation}
  \eta =   \frac{-4}{N\tilde{V}_{pp}}\sum_{\alpha, \kk,s}\!n_{\rm F}(\xi_{\alpha\kk s})\frac{\partial \xi_{\alpha\kk s}}{\partial \eta}.
  \label{eq:selfconeta}
\end{equation}

Second-order phase boundaries for nematic ordering can be determined using the stability condition of $\omega$
by requiring
\begin{multline}
  \left.\frac{\partial^2\omega}{\partial\eta^2}\right|_{\eta=0} =
  \Big( \frac{\tilde{V}_{pp}}{4}-\frac{1}{N}\sum_{\alpha, \kk, s}\frac{1}{4T\cosh^2\frac{\xi_{\alpha\kk s}}{2T}}\Big(\frac{\partial\xi_{\alpha \kk s}}{\partial \eta}\Big)^2 \\
  + \left.  \frac{1}{N}\sum_{\alpha, \kk, s} n_{\rm F}(\xi_{\alpha\kk
      s})\frac{\partial^2\xi_{\alpha \kk s}}{\partial \eta^2} \Big)\right|_{\eta=0}=0.
  \label{eq:omegaetaeta}
\end{multline}
This is equivalent to analyzing the linearized self-consistency
equation for $\eta$. To additionally find first-order phase boundaries requires
examining the grand potential for the global minimum.

\subsubsection{Nematic-spin-nematic order}
In complete analogy to the mean-field decoupling introduced above for the nematic order, we find the mean-field Hamiltonian for the nematic-spin-nematic ordering with non-vanishing $\eta_s$ as defined in Eq.~\eqref{eq:etas}. We only have to replace the diagonal elements of the Hamiltonian \eqref{eq:h0k} by the (now spin-dependent)
\begin{eqnarray}
  \xi_{xs} &=& \Delta + \tilde{U}_{p} \frac{n^{p}}{4} - s U_{p}\frac{\eta_s}{4} - \mu,\\
  \xi_{ys} &=& \Delta + \tilde{U}_{p} \frac{n^{p}}{4} + s U_{p}\frac{\eta_s}{4} - \mu,\\
  \xi_{ds} &=& \tilde{U}_{d} \frac{(n-n^p)}{2} - \mu,
  \label{eqn:diagnsn}
\end{eqnarray}
and
\begin{equation}
  \frac{f(n^{p}, \eta_s)}{N} = -\tilde{U}_{p} \frac{(n^{p})^2}{8} + U_{p}\frac{\eta_s^{2}}{8} - \tilde{U}_{d}\frac{(n-n^{p})^2}{4}.
  \label{eq:fnsn}
\end{equation}
The interaction parameters $\tilde{U}_p$ and $\tilde{U}_{d}$ are again given by Eqs.~\eqref{eq:Uptilde} and \eqref{eq:Udtilde}, respectively.
Note that now, the interaction driving the instability is not the
oxygen-oxygen nn interaction $V_{pp}$, but the oxygen on-site
interaction $U_p$. This ordering is thus in direct competition with an overall magnetization $m$ on the oxygen sites as given in Eq.~\eqref{eq:mag}. The critical $U_p$ for a nematic-spin-nematic  instability to occur is again determined by analyzing the stability condition for the corresponding grand potential, $\partial^2_{\eta_s} \omega = 0$.

\subsection{Results}
\label{subsec:results}
\begin{figure}[tb]
  \begin{center}
    \includegraphics{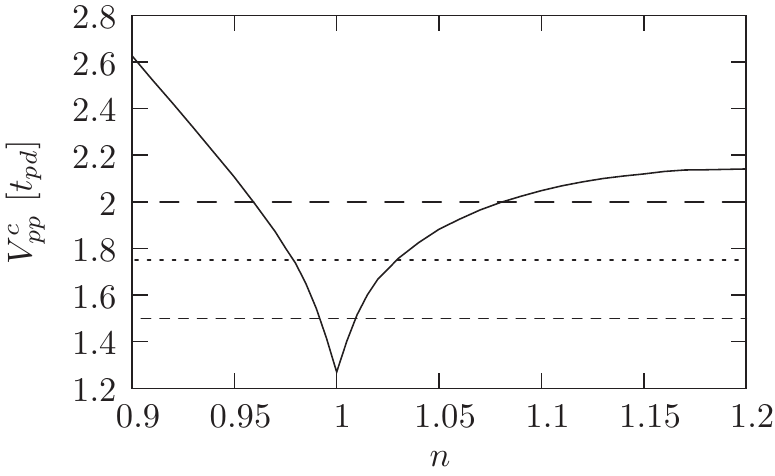}
  \end{center}
  \caption{The critical strength of the oxygen-oxygen interaction
    $V_{pp}^c$ needed in order to enter a nematic phase as a function
    of hole density $n$ for $t_{pp}=0$. For numerical reasons, the
    calculation has been carried out at $T=5\cdot10^{-4}[t_{pd}]$. The
    dashed lines denote the values of $V_{pp}$ used for
    Fig.~\ref{fig:tcvpp}.}
  \label{fig:vppc}
\end{figure}
\subsubsection{Nematic order}
Our goal is to investigate the effect each parameter has on the
nematic instability. 
For this, we use the linearized self-consistency
equation~\eqref{eq:omegaetaeta}
to map out various phase boundaries in the parameter space. 
Due to the large parameter space of the three-band model
we
present results with 
$\Delta=2.5$, $U_{p} = 3$ and $V_{pd}=1$ as realistic values for the
cuprates. 
Realistic values for the O-O hopping and the Cu on-site interaction
are $t_{pp}\approx 0.2$-$0.5$ and $U_{d}\approx
6$-$8$.\cite{mcmahan:90, hybertsen:89} After calculating a general phase diagram and looking at the influence of finite O-O hopping, we can thus for example analyze the effect of the Cu on-site interaction on the nematic phase formation.

IUC nematic ordering within this mean-field theory arises through a Stoner-type instability. It therefore requires a finite
interaction strength for all hole densities away from the van
Hove filling, where the diverging density of states allows for an
instability at infinitesimal $V_{pp}$.
Fig.~\ref{fig:vppc} shows the critical oxygen-oxygen
interaction strength  $V_{pp}^c$ needed to enter a nematic phase at
$T\approx0$ for  Cu on-site interaction strength $U_d = 9$ and
$t_{pp}=0$ with the van Hove singularity at $n_{\rm vH}=1$.
Note that the doping dependence of $V_{pp}^c$ is not symmetric around
the van Hove point. This is a multi-band effect and we will return to this in
section~\ref{subsec:oneband}.

\begin{figure}[bt]
  \begin{center}
    \includegraphics{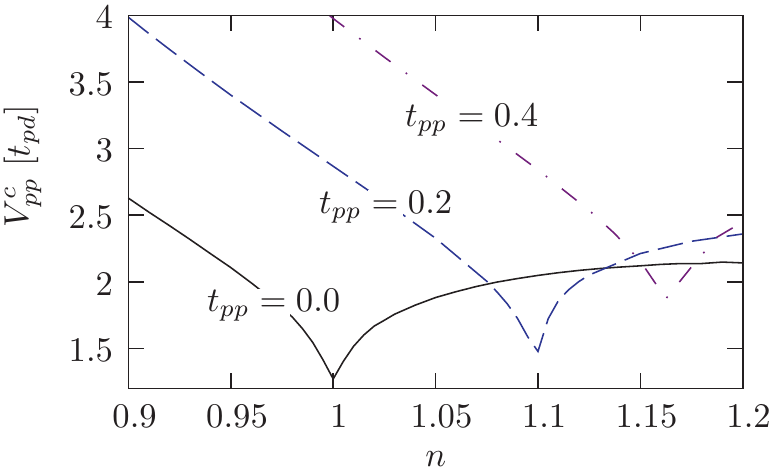}
  \end{center}
  \caption{Critical interaction strength for different values of the oxygen-oxygen hopping $t_{pp}$.}
  \label{fig:vctpp}
\end{figure}

The main effect of a finite oxygen-oxygen hopping $t_{pp}$ is to shift
the van Hove singularity to higher hole densities as can be seen in
Fig.~\ref{fig:vctpp}, where
we show the hole-density dependence of $V_{pp}^c$
for different $t_{pp}$.
In addition
we see that the nematicity is gradually suppressed upon an increase in
$t_{pp}$,
which reflects the fact that a finite $t_{pp}$ adds to the
 two-dimensionality of the system.
For the rest of this section, we will focus on the case of $t_{pp} =
0$.
\footnote{For $\eta\rightarrow 0$, we can then treat the nematic part of the Hamiltonian as a perturbation.  This is especially useful for evaluating the linearized self-consistency equation~\eqref{eq:omegaetaeta}, which is done at $\eta = 0$. The details of this perturbation expansion are shown in App.~\ref{app:tpp0}}
\begin{figure}[bt]
  \begin{center}
    \includegraphics{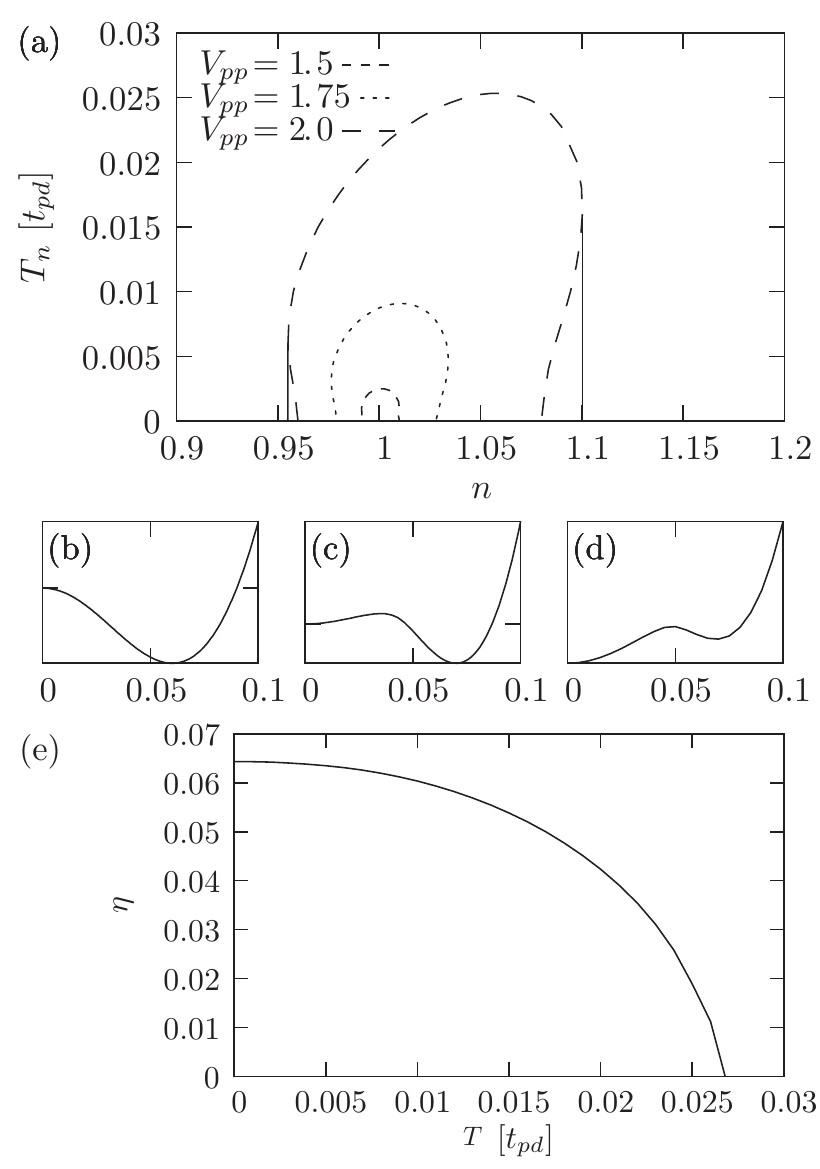}
  \end{center}
  \caption{(a) Phase diagram for the different values of the O-O
    nearest-neighbor interaction $V_{pp} = 2$, $1.75$ and $1.5$. At
    low temperature, there would be first-order transitions, only
    shown for $V_{pp} = 2$ by the solid lines, before the normal state
    becomes unstable (dashed lines). Figs.~(b) - (d) show the free energy
    as a function of $\eta$ for $n=1.05$, $n=1.095$ and $n=1.105$ at $T=0.001$,
    illustrating the first-order character of the low temperature
    transition. 
    (e) The nematic order parameter $\eta$ as a function of
    temperature for $U_{d}=9$ and $n=1.05$ showing the second-order
    transition at $T_c=0.027$.}
  \label{fig:tcvpp}
\end{figure}

We now turn to the $T$-$n$ phase diagram shown in Fig.~\ref{fig:tcvpp}(a).
For the phase diagram, we obtain the second-order
phase boundary from the linearized self-consistency equation
for $V_{pp} = 2$, $1.75$ and $1.5$ (dashed horizontal lines in Fig.~\ref{fig:vppc}). For $V_{pp}=1.5$, a small dome almost symmetric around the van
Hove filling is found while for higher $V_{pp}$, the dome becomes
asymmetric with respect to the van Hove filling $n_{\rm vH}=1$.
At low temperature, we expect the phase transition to be of first order as was shown for one-band models within mean-field theory in Ref.\cite{yamase:2005}.
Explicitly examining the full grand potential for $V_{pp}$, we indeed find second-order transitions at higher temperature,
i.e. for hole densities in the middle of the dome and first-order
transitions for densities at the border of the dome, as indicated for
$V_{pp}=2$ by the solid lines in Fig.~\ref{fig:tcvpp}(a). Note that in the case of a one-band description it was shown that fluctuations may make the first-order transitions continuous~\cite{jakubczyk:2009} and we thus expect a similar behaviour here. 
 
To illustrate the first-order character of the low-temperature transition, Figs.~\ref{fig:tcvpp}(b)-(d) show the free energy as a function of $\eta$ at $T\approx 0$ for a hole density $n$ deep inside the phase, where the normal state is metastable, and where the nematic state is metastable. The second-order character of the transition on the top of the dome is best seen in the $T$ dependence of the order parameter $\eta$. This dependence is shown in Fig.~\ref{fig:tcvpp}(e).  

\begin{figure}[tb]
  \begin{center}
    \includegraphics{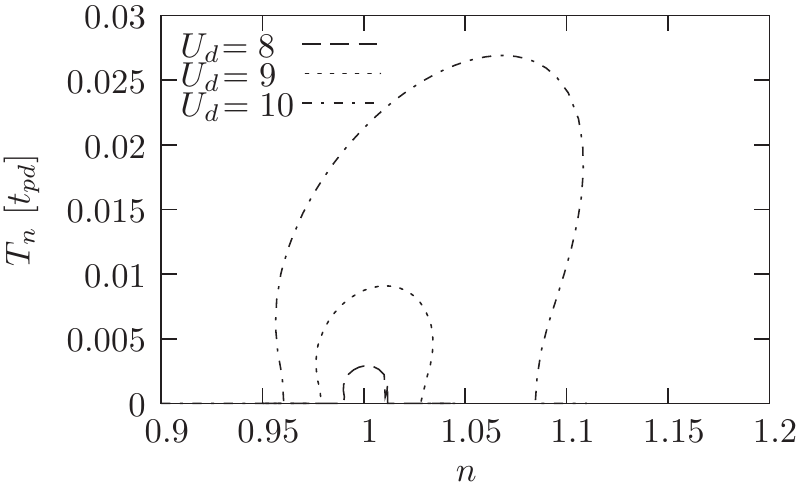}
  \end{center}
  \caption{Phase diagram for $V_{pp} = 1.75$ and different Cu on-site interaction strengths $U_{d}=8, 9, 10$. Shown are again only `second-order' phase boundaries.}
  \label{fig:tcud}
\end{figure}

What is particularly noteworthy from our survey of parameter space is
that the Cu on-site interaction $U_d$ tends to stabilizes the
nematic phase as shown in Fig.~\ref{fig:tcud}.
In fact we find that the effect of
increasing $U_d$ is almost the same as increasing $V_{pp}$ as is
apparent upon comparison of Figs.~\ref{fig:tcvpp}(a) and
\ref{fig:tcud}.
As discussed in section \ref{subsec:oneband},
this is due to an increased hole density at the oxygen sites as well as a reduced level separation between the lowest lying bands for larger $U_d$.

\subsubsection{Nematic-spin-nematic order}
\begin{figure}[tb]
  \begin{center}
    \includegraphics{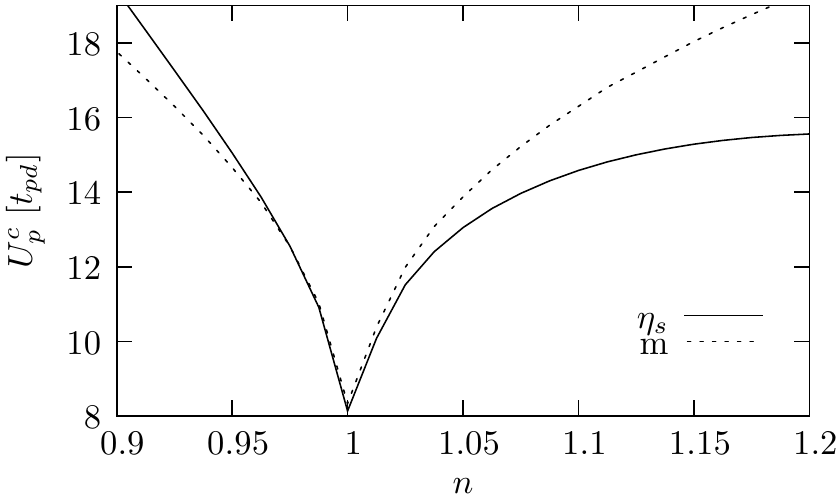}
  \end{center}
  \caption{Critical oxygen interaction strength for a nematic-spin-nematic ($\eta_{s}$) and a magnetic ($m$) instability on the oxygen sites. Here, $U_{d} = 9$, $V_{pd} = V_{pp} = 1$ and $t_{pp} = 0$. }
  \label{fig:Upc}
\end{figure}

We only present
the doping
dependence of the critical
oxygen on-site interactions $U_p^c$ which drives the
nematic-spin-nematic (see  
Fig.~\ref{fig:Upc}), as technical details of 
the self-consistency analysis for the nematic-spin-nematic phase are
very much analogous to that for the nematic phase.
For
completeness, we also show the critical $U_p$ for a
ferromagnetic instability with $m$ as defined in Eq.~\eqref{eq:mag} 
and $\eta_s=0$. This competing ferromagnetic instability is only
favored over the nematic-spin-nematic order for $n<1$ (electron doping).
Notice  that
the magnitude of  $U_p^c$ 
for the nematic-spin-nematic order to occur is almost an order of
magnitude larger than the magnitude of $V_{pp}^c$
for the nematic instability. 
This is due to the fact that here, the holes on an oxygen site with spin
$s$ are only interacting with the holes on the same site with opposite
spin, while for the nematic instability, the holes interact with the total
hole density of all four neighboring oxygen sites.

Now we can compare the influence of various interaction strengths for
nematic-spin-nematic ordering to that  for nematic ordering. We find
that $U_d$, $\Delta$ and $t_{pp}$ have the same effect for both types
of ordering: 
$U_d$ increasing and $\Delta$ and $t_{pp}$ decreasing the tendency towards
both orders. However,  increase in $V_{pp}$ leads to larger $U_{p}^c$ as it
reduces hole occupation of the oxygen sites.
We thus find that the interaction driving the nematic or nematic-spin-nematic instability hurts the occurrence of the respective other phase.

\subsection{Comparison to one-band model}
\label{subsec:oneband}
 In this subsection, we highlight similarities and differences between the mean-field
theory of the three-band Emery model and previous studies of effective
one-band models. We first derive effective attractive interactions in
spin-symmetric and  antisymmetric channels with $d_{x^2-y^2}$ symmetry
for the lowest band of non-interacting model, in terms of {\em repulsive}
oxygen-oxygen interactions. We then discuss the multi-band effect in
the ``Stoner-like'' self-consistency condition. 

In order to see how $V_{pp}$ and $U_p$ lead to effective attractive
interactions for the lowest lying band $\xi_{1\kk
  s}^{0}$,  we express the interaction term
in the diagonal basis $\hat{c}_{\alpha\kk s}$ of the non-interacting Hamiltonian (setting $t_{pp} = 0$). 
In this basis
the oxygen operators read (see App.~\ref{app:tpp0})
\begin{eqnarray}
  \hat{p}_{x \kk s} &=&   - \tilde{\gamma}_{1x}v_{\kk}\hat{c}_{1\kk s}-\tilde{\gamma}_{1y} \hat{c}_{2\kk s} + \tilde{\gamma}_{1x}u_{\kk}\hat{c}_{3\kk s},\\
  \hat{p}_{y \kk s} &=&  - \tilde{\gamma}_{1y}v_{\kk}\hat{c}_{1\kk s} -\tilde{\gamma}_{1x} \hat{c}_{2\kk s}+ \tilde{\gamma}_{1y}u_{\kk}\hat{c}_{3\kk s},
  \label{eq:newbasis}
\end{eqnarray}
where $u_{\kk}=\cos\frac{\omega_{\kk}}2$, $v_{\kk}=\sin\frac{\omega_{\kk}}2$ with
\begin{equation}
  \omega_{\kk} = \arctan\left(\frac{2\sqrt{\gamma_{1}^{2}(k_x) + \gamma_{1}^{2}(k_y)}}{\Delta}\right),
  \label{eq:omega}
\end{equation}
and 
\begin{equation}
  \tilde{\gamma}_{1i} =
  \frac{\gamma_{1}(k_{i})}{\sqrt{\gamma_{1}^2(k_x) +
      \gamma_{1}^{2}(k_y)}},\quad i=x,y
  \label{eq:gammatilde}
\end{equation}
with $\gamma_1 (k_{i})=-2t_{pd}\cos\frac{k_i}{2}$ as it was  defined in Eq.\eqref{eq:gammas}.
Note that the oxygen on-site interaction can be separated into spin-symmetric and antisymmetric parts as \begin{multline}
  \frac{U_{p}}{N} \sum_{\kk,\kk'}\sum_{\nu = x, y}\hat{n}_{\nu\kk \uparrow}\hat{n}_{\nu\kk'\downarrow}\\
  = \frac{U_{p}}{2N} \sum_{\kk,\kk'}\sum_{\nu = x, y}\sum_{s, s'}(\hat{n}_{\nu\kk s}\hat{n}_{\nu\kk' s'} - s s' \hat{n}_{\nu\kk s}\hat{n}_{\nu\kk' s'}),
  \label{eq:on-site2}
\end{multline}
while the oxygen nn interaction only has a spin-symmetric part.
We now start by expressing the 
spin-symmetric part of the inter-oxygen
interactions in the basis
$\hat{c}_{\alpha {\bf k}s}$. The terms that only involve lowest
bands are 
\begin{multline}
  \frac{U_{p}}{4N} \sum_{\kk,\kk'}\sum_{s, s'}(\tilde{\gamma}_{1x}^2\tilde{\gamma}_{1x'}^2 + \tilde{\gamma}_{1y}^2\tilde{\gamma}_{1y'}^2)v_{\kk}^2v_{\kk'}^2 \hat{n}_{1\kk s} \hat{n}_{1\kk' s'}\\
  + \frac{2V_{pp}}{N} \sum_{\kk,\kk'}\sum_{s, s'}(\tilde{\gamma}_{1x}^2\tilde{\gamma}_{1y'}^2 + \tilde{\gamma}_{1y}^2\tilde{\gamma}_{1x'}^2)v_{\kk}^2v_{\kk'}^2 \hat{n}_{1\kk s} \hat{n}_{1\kk' s'}
  \label{eq:both1}
\end{multline}
with $\hat{n}_{1\kk s} = \hat{c}_{1 \kk s}^{\dag}\hat{c}_{1\kk
  s}^{\phantom{\dag}}$  the density operator for the lowest lying
band.
Finally Eq.\eqref{eq:both1} can be reorganized into an isotropic
part
\begin{equation}  
  \frac{\tilde{U}_{p}}{4N} \sum_{\kk, \kk'}\sum_{s,
    s'}v_{\kk}^2v_{\kk'}^2\hat{n}_{1\kk s} \hat{n}_{1\kk' s'}
  \label{eq:on-site3}
  \end{equation}
where $\tilde{U}_p = (U_p + 8V_{pp})/2$, and the effective ``$F_{2,s}$'' part
\begin{equation}
  -\frac{\tilde{V}_{pp}}{4 N} \sum_{\kk, \kk'}\sum_{s, s'}d_{\kk}d_{\kk'}v_{\kk}^2v_{\kk'}^2\hat{n}_{1\kk s} \hat{n}_{1\kk' s'}
  \label{eq:nn3}
  \end{equation}
where $\tilde{V}_{pp} = (8V_{pp} - U_p)/2$ and
\begin{equation}
  d_{\kk} = \frac{(\cos k_x - \cos k_y)}{(2 + \cos k_x + \cos k_y)}.
  \label{eq:dk}
\end{equation}

Some remarks are in order. Eq.\eqref{eq:nn3} explicitly shows that
repulsive $V_{pp}$ leads to effective attractive interaction that can
drive nematicity for the lowest lying band. It also shows that $U_p$
hinders nematic ordering. Furthermore, we see that
these inter-oxygen interactions are acting only on 
the portion of hole density in the lowest lying band $\xi_{1\kk
  s}^{0}$ that can be attributed to oxygen
occupation since 
the oxygen occupation number
\begin{equation}
  n^p = \frac{1}{N}\sum_{\kk, s}v_{\kk}^2\, n_{\rm F}(\xi_{1\kk s}^{0}).
  \label{eq:npeta0}
\end{equation}

Following the same procedure, the spin-antisymmetric part of
Eq.\eqref{eq:on-site2}  can be organized into an
isotropic part and the effective ``$F_{2,a}$'' part:
\begin{multline}
  -\frac{U_{p}}{8N} \sum_{\kk, \kk'}\sum_{s, s'} s s' v_{\kk}^2v_{\kk'}^2\hat{n}_{1\kk s} \hat{n}_{1\kk' s'}\\
  -\frac{U_p}{8 N} \sum_{\kk, \kk'}\sum_{s, s'}s s' d_{\kk}d_{\kk'}v_{\kk}^2v_{\kk'}^2\hat{n}_{1\kk s} \hat{n}_{1\kk' s'}.
  \label{eq:nsn_m}
\end{multline}
We therefore find explicitly that oxygen nearest-neighbor and on-site
interactions in the three-band model lead in a one-band model to
attractive spin-symmetric and antisymmetric interactions of
$d_{x^2-y^2}$ symmetry, $F_{2,s}$ and $F_{2,a}$, driving nematic and
nematic-spin-nematic order, respectively.

We now turn to the multi-band effect in  
the linearized self-consistency equation~\eqref{eq:omegaetaeta}.
For the parameter space of interest to the cuprates,
only the lowest band of the (mean-field) Hamiltonian
with energy $\xi_{1\kk s}$ is filled
at low temperatures. Hence, Eq.~\eqref{eq:omegaetaeta} amounts to
\begin{multline}
  \Big(\frac{\tilde{V}_{pp}}{4} - \frac{1}{N}\sum_{\kk, s}\frac{1}{4T\cosh^2\frac{\xi_{1\kk s}}{2T}}\Big(\frac{\partial\xi_{1 \kk s}}{\partial \eta}\Big)^2\\
  \left. + \frac{1}{N}\sum_{\kk, s} n_{\rm F}(\xi_{1\kk s})\frac{\partial^2\xi_{1 \kk s}}{\partial \eta^2}\Big)\right|_{\eta=0} = 0.
  \label{eq:linself}
\end{multline}
While the first line is the familiar result from one-band mean-field calculations with the second term being the familiar polarization bubble, the term on the second line has no analogue in simple single-band models. This term grows with total hole density and is thus responsible for the asymmetry around the van Hove filling found in Sec.~\ref{subsec:results}.
\begin{figure}[tb]
  \centering
  \includegraphics{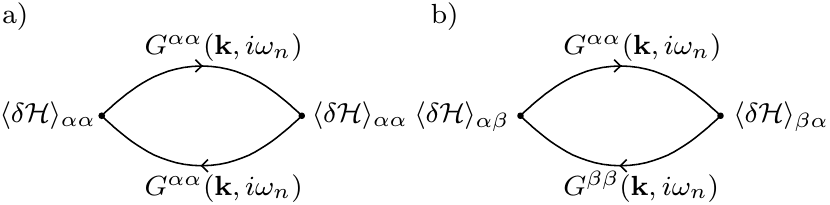}
  \caption{The two bubble diagrams involved in the linearized self-consistency equation, where we have used the short notation $\langle \delta\HH\rangle_{\alpha\beta} = \langle \alpha|\delta\HH|\beta\rangle$.}
  \label{fig:bubbles}
\end{figure}

To better understand Eq.~\eqref{eq:linself}, we interpret nematic order as a perturbation around the isotropic ($\eta=0$) Hamiltonian $\HH_{\kk s}^{({\rm iso})}$ (as done in App.~\ref{app:tpp0} for $t_{pp}=0$) and write the derivatives in Eq.~\eqref{eq:linself} in terms of the perturbation-theory expansion parameters, see Eqs.~\eqref{eq:xinp} - \eqref{eq:xietaeta}. Introducing the (isotropic) Green's functions $G_{\alpha}^{({\rm iso})}(\kk, \omega_n) = (i\omega_n - \xi^{({\rm iso})}_{\alpha \kk s})^{-1}$ with $\xi^{({\rm iso})}_{\alpha \kk s} = \xi_{\alpha \kk s}|_{\eta=0}$ the eigenenergies of $\HH_{\kk s}^{({\rm iso})}$ and using the relation
\begin{multline}
  \left.\sum_{\omega_{n}}G^{({\rm iso})}_{\alpha}(\kk, \omega_n)G^{({\rm iso})}_{\alpha}(\kk + \q, \omega_n)\right|_{\q \rightarrow 0}\\
  =\left. \frac{n_{\rm F}(\xi^{({\rm iso})}_{\alpha\kk s}) - n_{\rm F}(\xi^{({\rm iso})}_{\alpha\kk+\q s})}{\xi^{({\rm iso})}_{\alpha\kk s} - \xi^{({\rm iso})}_{\alpha\kk + \q s}}\right|_{\q \rightarrow 0} = \frac{\partial n_{\rm F}(\xi^{({\rm iso})}_{\alpha \kk s})}{\partial \xi},
 \label{eq:Gsums}
\end{multline}
we find
\begin{multline}
  \left.\frac{1}{N}\sum_{\kk, s}\frac{1}{4T\cosh^2\frac{\xi_{1\kk s}}{2T}}\Big(\frac{\partial\xi_{1 \kk s}}{\partial \eta}\Big)^2\right|_{\eta = 0}\\
  = \frac{1}{N}\sum_{\kk s} G^{({\rm iso})}_{1}(\kk, \omega_n)G^{({\rm iso})}_{1}(\kk, \omega_n)\langle1|\delta \HH|1\rangle,
 \label{eq:firstterm}
\end{multline}
the familiar polarization bubble as depicted in Fig.~\ref{fig:bubbles}(a). For the case of multiple bands, also inter-band interactions should be taken into account as indicated by Fig.~\ref{fig:bubbles}(b). The corresponding expression yields
\begin{multline}
  \sum_{\alpha\neq\beta}\sum_{\kk,\omega_n}G^{({\rm iso})}_{\alpha}(\kk, \omega_n)G^{({\rm iso})}_{\beta}(\kk, \omega_n)|\langle\alpha|\delta\HH|\beta\rangle|^2\\
  = \sum_{\alpha\neq\beta}\sum_{\kk}\left[n_{\rm F}(\xi^{({\rm iso})}_{\alpha\kk s}) - n_{\rm F}(\xi^{({\rm iso})}_{\beta\kk s})\right]\frac{|\langle\alpha|\delta\HH|\beta\rangle|^2}{\xi^{({\rm iso})}_{\alpha\kk s} - \xi^{({\rm iso})}_{\beta\kk s}}\\
  = \left.\frac{1}{N}\sum_{\kk, s} n_{\rm F}(\xi_{1\kk s})\frac{\partial^2\xi_{1 \kk s}}{\partial \eta^2}\right|_{\eta = 0},
 \label{eq:sum2}
\end{multline}
where in the last step we have used Eq.~\eqref{eq:xietaeta} and the fact, that only the lowest band is occupied for low temperatures. The asymmetry is thus a multi-band effect unlike the asymmetry found in Ref.~\cite{yamase:2005}, which is due to an asymmetric density of states. It is now clear why the asymmetry only appears for large values of $V_{pp}$: only when the O-O nn interaction is comparable to the band separation, the influence of this term becomes visible. Increasing $U_d$ has then two effects, both enhancing nematicity: in addition to increasing the hole density on the oxygen sites, it shifts the lowest band slightly up in energy, decreasing the level separation to the second band, hence increasing the importance of the inter-band term. This explains why an increase in $U_d$ has such a similar effect as an increase in $V_{pp}$.

\section{loop currents}
\label{sec:loop}
\begin{figure}[tb]
  \begin{center}
    \includegraphics{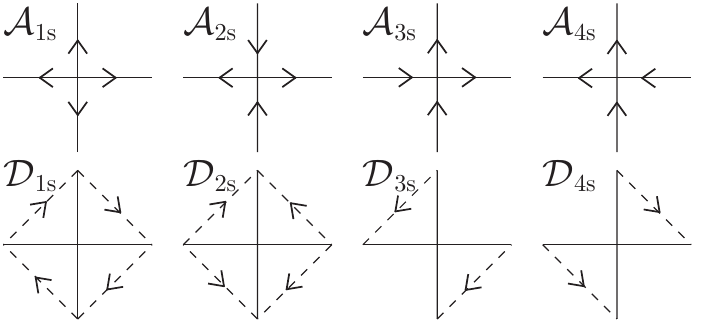}
  \end{center}
  \caption{The different current patterns arising from the operators $\mathcal{A}_{1-4 s}$ in Eqs.~\eqref{eq:a12} and \eqref{eq:a34} and $\mathcal{D}_{1-4 s}$ in Eqs.~\eqref{eq:d12} - \eqref{eq:d4}. Combining $\mathcal{A}_{2s}$ with $\mathcal{D}_{2s}$ leads to the loop-current phase $\Theta_{I}$, while $\mathcal{A}_{3s}$ ($\mathcal{A}_{4s}$) combined with $\mathcal{D}_{3s}$ ($\mathcal{D}_{4s}$) leads to $\Theta_{II}$.}
  \label{fig:loops}
\end{figure}
In the loop-current picture, the staggered magnetic moments observed in experiments~\cite{fauque:2006} originate in circulating electron currents around O-Cu-O triangles within the unit cell. This phase was introduced and analyzed in a mean-field approach by Varma.~\cite{varma:2006b} Stipulating a Cu-O hopping integral that depends on the hole doping and a vanishing charge-transfer gap, a phase-diagram was drawn in qualitative agreement with the pseudogap phase. 
While later exact diagonalization calculation on small clusters concluded that the energy scale of such current loops was too small to account for the phenomena associated with the pseudogap region,~\cite{greiter:2008} RG calculations for two-leg ladders found incommensurate loop currents\cite{chudzinski:2007} and a variational Monte Carlo study found that the $\Theta_{II}$ current pattern is stabilized in intermediate system sizes\cite{weber:2009}.

In this section, our aim is not to answer the question whether such loop currents exist in the parameter range usually assumed for the cuprates, but again to investigate the microscopic origin and the influence of the various model parameters. This then allows for a comparison with the two orderings of the previous section. 

\subsection{Mean-field theory of loop currents}
 To analyze this loop phase, we perform a similar calculation as in
 Ref.~\cite{varma:2006b}, however with some important differences:
 First, we only use the Hamiltonian as defined in
 Eqs.~\eqref{eq:hopping} and \eqref{eq:interaction} without any
 implicit assumption regarding doping dependence of parameters.
 Further, our calculation includes the O-O nn interaction, which we decouple analogously to the Cu-O nn interaction.	   
 Finally, we do not reformulate the Hamiltonian in terms of flux
 through the Cu-O triangles, but decouple the nn interaction terms and
 directly solve for the self-consistent mean-field solution. Our
 conventional treatment of the problem leads to different doping
 dependence and an additional $V_{pp}$ contribution compared to Ref. ~\cite{varma:2006b}.

We start with the interactions in Eq.~\eqref{eq:interaction} and follow the Cu-O-interaction decoupling of Varma~\cite{varma:2006b} by defining the operators
\begin{multline}
  \mathcal{A}^{\dag}_{1/2 is} = \frac12\Big[(\hat{d}^{\dag}_{i,s}\hat{p}^{\phantom{\dag}}_{i+\hat{x}/2, s}+\hat{d}^{\dag}_{i,s}\hat{p}^{\phantom{\dag}}_{i-\hat{x}/2, s})\\ 
  \pm (\hat{d}^{\dag}_{i,s}\hat{p}^{\phantom{\dag}}_{i+\hat{y}/2, s}+\hat{d}^{\dag}_{i,s}\hat{p}^{\phantom{\dag}}_{i-\hat{y}/2, s})\Big],
  \label{eq:a12}
\end{multline}
\begin{multline}
  \mathcal{A}^{\dag}_{3/4 i s} = \frac i2\Big[(\hat{d}^{\dag}_{i,s}\hat{p}^{\phantom{\dag}}_{i+\hat{x}/2, s}-\hat{d}^{\dag}_{i,s}\hat{p}^{\phantom{\dag}}_{i-\hat{x}/2, s})\\
  \pm (\hat{d}^{\dag}_{i,s}\hat{p}^{\phantom{\dag}}_{i+\hat{y}/2, s}-\hat{d}^{\dag}_{i,s}\hat{p}^{\phantom{\dag}}_{i-\hat{y}/2, s})\Big].
  \label{eq:a34}
\end{multline}
Introducing the (complex) mean-field order parameters
\begin{equation}
  R_{\nu} e^{i \phi_{\nu}} = V_{pd}\sum_{s}\langle \mathcal{A}_{\nu s}\rangle
  \label{eq:mftop}
\end{equation}
with $\langle \mathcal{A}_{\nu s}\rangle = \langle \mathcal{A}_{\nu i s}\rangle$ independent of site $i$, the Cu-O interaction can straight-forwardly be mean-field decoupled in the Fock channel
\begin{multline}
  -V_{pd}\sum_{i,\nu}\sum_{s, s'} \mathcal{A}^{\dag}_{\nu i s} \mathcal{A}^{\phantom{\dag}}_{\nu i s'}\approx\\
  -\sum_{i, \nu, s} (R_{\nu}e^{-i\phi_{\nu}} \mathcal{A}^{\phantom{\dag}}_{\nu i s} + {\rm h.c.}) + \frac{N}{V_{pd}} \sum_{\nu}R_{\nu}^2.
  \label{eq:newvpd}
\end{multline}
The order parameters $R_{\nu}$ correspond to the current patterns shown in Fig.~\ref{fig:loops} and can again be classified according to the symmetries they break. We first note that $\mathcal{A}_{1 s}$ can not lead to a stationary current loop. Focussing for the other order parameters again only on $C_4$, $\mathcal{I}$, $\mathcal{T}$ and combinations thereof, we find that the order parameter arising from $\mathcal{A}_{2s}$, corresponding to $\Theta_I$ in Ref.~\cite{varma:2006b}, differs from the order parameters arising from $\mathcal{A}_{3s}$ and $\mathcal{A}_{4s}$ corresponding to the $\Theta_{II}$ phase, in that it preserves $C_4\circ \mathcal{T}$ (see Tab.~\ref{tab:sym}). 

Next, we similarly look at the O-O interaction. For the decoupling, we again introduce operators of distinct symmetry,
\begin{multline}
  \mathcal{D}^\dag_{1/2 i s} = \frac{1}{\sqrt2}\Big(\hat{p}^\dag_{i-\hat{x}/2, s} \hat{p}^{\phantom{\dag}}_{i+\hat{y}/2, s} \mp \hat{p}^\dag_{i+\hat{x}/2, s} \hat{p}^{\phantom{\dag}}_{i+\hat{y}/2, s}\\
  + \hat{p}^\dag_{i+\hat{x}/2, s} \hat{p}^{\phantom{\dag}}_{i-\hat{y}/2, s} \mp \hat{p}^\dag_{i-\hat{x}/2, s} \hat{p}^{\phantom{\dag}}_{i-\hat{y}/2, s}\Big)
  \label{eq:d12}
\end{multline}
and 
\begin{eqnarray}
  \mathcal{D}^\dag_{3 i s} &=& i\Big(\hat{p}^\dag_{i+\hat{x}/2, s} \hat{p}^{\phantom{\dag}}_{i+\hat{y}/2, s} - \hat{p}^\dag_{i-\hat{x}/2, s} \hat{p}^{\phantom{\dag}}_{i-\hat{y}/2, s}\Big)\label{eq:d3},\\
  \mathcal{D}^\dag_{4 i s} &=& i\Big(\hat{p}^\dag_{i-\hat{x}/2, s} \hat{p}^{\phantom{\dag}}_{i+\hat{y}/2 ,s} - \hat{p}^\dag_{i+\hat{x}/2, s} \hat{p}^{\phantom{\dag}}_{i-\hat{y}/2, s}\Big).
  \label{eq:d4}
\end{eqnarray}
This allows us to introduce (site-independent) mean-fields
\begin{equation}
  R^{p}_{\nu}e^{i\phi^p_{\nu}} = V_{pp}\sum_s \langle \mathcal{D}_{\nu s}\rangle
  \label{eq:oomf}
\end{equation}
and decouple the O-O interaction term in the Fock channel as
\begin{multline}
  -\frac{V_{pp}}2 \sum_{i, \nu} \sum_{s, s'} \mathcal{D}^{\dag}_{\nu i s} \mathcal{D}^{\phantom{\dag}}_{\nu i s'}\approx \\
  -\frac{1}2 \sum_{i,\nu, s}(R^p_{\nu} e^{-i\phi^p_{\nu}}\mathcal{D}_{\nu i s}+ {\rm h.c.}) + \frac{N}{2V_{pp}}\sum_{\nu}(R^p_{\nu})^2.  
  \label{eq:dddecoupling}
\end{multline}
Looking at Fig.~\ref{fig:loops}, we see that $\mathcal{D}_{1s}$ only breaks time-reversal symmetry, while $\mathcal{D}_{2s}$ leads to the $\Theta_I$ phase and $\mathcal{D}_{3/4 s}$ to the $\Theta_{II}$ phase with the respective broken symmetries. 

In the following, we are only interested in the phase $\Theta_{II}$
and therefore only keep the two order parameters $R_{3}\equiv R$ and
$R^p_{3}\equiv R_p$, which have the same symmetry and mix,
finite. Analogously, we could also choose $R_{4}$ and $R^p_{4}$ (see
Fig.~\ref{fig:loops}). An order parameter yielding a current
(rather than a bond-density) has to have an imaginary part and for
simplicity, we set the phases to $\phi_3=\phi^p_3 =
\pi/2$.\footnote{For the case, where only the Cu-O interaction is
  considered, Varma showed that this choice indeed minimizes the
  Energy.\cite{varma:2006b}} Note that non-zero $R$ and $R_p$, while
corresponding to non-zero loop-currents, do not guarantee the absence
of macroscopic currents.

Contrary to Sec.~\ref{sec:static}, where the mean-field decoupling led to a shift of the diagonal elements in the Hamiltonian, here it leads to new hopping elements in the Hamiltonian~\eqref{eq:h0k}
\begin{equation}
  \tilde{\gamma}_1(k_i) = -2t_{pd}\cos\frac{k_i}2 - i R \sin\frac{k_i}2,
  \label{eq:currentgamma1}
\end{equation}
and
\begin{multline}
  \tilde{\gamma}_2(\kk) = -4 t_{pp} \cos\frac{k_x}2\cos\frac{k_y}2\\
  - i R_p (\sin\frac{k_x}2\cos\frac{k_y}2 - \cos\frac{k_x}2\sin\frac{k_y}2).
  \label{eq:currentgamma2}
\end{multline}
For simplicity, we decouple the nearest-neighbor interaction only in the loop-current (Fock) channel and thus the diagonal elements yield
\begin{eqnarray}
  \xi_{x,y} &=& \Delta + U_{p} \frac{n^{p}}{4} - \mu,\\
  \xi_{d} &=& U_{d} \frac{(n-n^p)}{2} - \mu,
  \label{eqn:currentdiag}
\end{eqnarray}
and
\begin{equation}
  \frac{f(n^{p},R, R_p)}{N} = - \frac{(n^p)^2}8 U_p - \frac{(n-n^p)^2}4 U_d + \frac{R^2}{V_{pd}} + \frac{R_p^2}{2 V_{pp}}.
  \label{eq:currentfN}
\end{equation}
To investigate the occurrence of instabilities, we need to account for the fact that the two order parameters $R$ and $R_p$ are coupled by symmetry. This means that we can not investigate their respective instabilities separately as done for the order parameters in the previous section. Instead, the pairs of critical interaction strengths $(V_{pp}^c, V_{pd}^c)$ are given by a vanishing eigenvalue of the Hessian matrix of the grand potential $\omega$,
\begin{equation}
  [\omega]_{\mu\nu} = \left(\begin{array}{cc}\partial_{R}^2\omega & \partial_{R}\partial_{R_p}\omega \\ \partial_{R_p}\partial_{R}\omega & \partial_{R_p}^2\omega\end{array}\right)\Big|_{R=R_p=0}.
  \label{eq:hessian}
\end{equation}

\begin{figure}[tb]
  \begin{center}
    \includegraphics{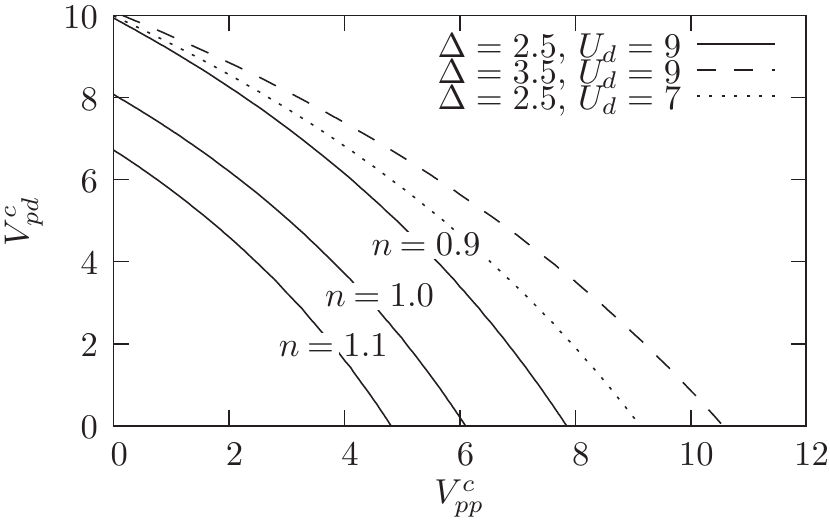}
  \end{center}
  \caption{Critical interactions $(V_{pp}^c, V_{pd}^c)$ for $U_d = 9$, $U_p = 3$, $t_{pp} = 0.1$, $\Delta = 2.5$ and different hole densities. The dashed and dotted lines for $n=0.9$ illustrate the influence of the Cu on-site interaction and the charge transfer gap.}
  \label{fig:varma}
\end{figure}

\subsection{Results}
The solid lines in Fig.~\ref{fig:varma} show the curves $(V_{pp}^c, V_{pd}^c)$ for $U_d= 9$, $U_p = 3$, $t_{pp} = 0.1$ and different hole densities. Due to coupling of the two order parameters $R$ and $R_p$, the critical Cu-O nn interaction $V_{pd}^c$ is reduced by a finite $V_{pp}$. We again study the influence of different parameters on the critical interaction values. As the dashed and the dotted lines for $n=0.9$ show, increasing the charge transfer gap $\Delta$ or reducing the copper on-site interaction $U_d$ results in higher critical interaction strengths. However, this mainly affects $V_{pp}^c$ due to the change in the oxygen hole occupancy, while $V_{pd}^{c}$ is almost unchanged. 

Contrary to the nematic and the nematic-spin-nematic order of the
previous section, the critical interaction strengths here are
monotonically decreasing with increasing hole density (see
Fig.~\ref{fig:varma}). This is due to the fact that the current
loop in a mean-field approach arises due to a Fock-type 
rather than Hartree-type decoupling and hence not 
a Stoner-type instability.
As the whole dispersion is altered by the decoupling, increasing the hole density in the lowest band increases the tendency towards loop currents. In order to find a phase diagram as found in the cuprates, additional assumptions to the model have to be made, such as a density-dependent hopping, e.g. of the form $t_{pd}\rightarrow t_{pd}|x|$ with $x = n-1$, as in Varma's analysis.~\cite{varma:2006b}

\section{discussion and conclusions}
\label{sec:discussion}
Starting from a three-band model and applying a mean-field approach - despite its obvious shortcomings - we gained valuable insights about the microscopic repulsive interactions that can promote various IUC orders.
We found that the Cu on-site interaction $U_{d}$ increases the tendency towards all the studied orderings by shifting more holes to the oxygens. The charge transfer gap $\Delta$ has the opposite effect.
Also, different interaction parameters affect the different instabilities differently: while the O on-site repulsion $U_{p}$ only favors the nematic-spin-nematic phase and the Cu-O repulsion $V_{pd}$ the loop currents, the nearest-neighbor O-O repulsion $V_{pp}$ helps both, the nematic and the loop-current phase (see Tab.~\ref{tab:params} for a summary of all the model parameters). 
Further, we could microscopically motivate attractive interactions $F_{2,s}$ and $F_{2,a}$ with a repulsive (longer-ranged) O-O repulsion $V_{pp}$ and O on-site repulsion $U_p$, respectively 

\begin{table}[b!]
  \centering
  \begin{tabular}{c|c|c|c|c|c|c}
    & $U_d$ & $U_p$ & $V_{pd}$ & $V_{pp}$ & $t_{pp}$ & $\Delta$\\
    \hline
    \hline
    nematic & + & - & - & + & - & -\\
    \hline
    nematic-spin-nematic & + & + & - & - & - & -\\
    \hline
    $\Theta_{II}$ loop current & + & -& + & + & - & -
  \end{tabular}
  \caption{Summary of the effect of the different parameters in the Emery model on the different IUC orders, where + denotes a parameter that helps a specific order and a - denotes a hindering parameter.}
  \label{tab:params}
\end{table}

A comment on the magnitude of the interactions necessary found here is in order: the energy scale of the pseudogap phase in the cuprates is of order 100 Kelvin. For any of the above phases to reach to such high temperatures, unrealistically large interactions are needed within our mean-field calculation. Looking at the nematic phase for example and taking $t_{pd}\approx1eV$, $V_{pp}$ needs to be of order $2t_{pd}$ as can be deduced from Fig.~\ref{fig:tcvpp}(a). Also, increasing the Cu on-site interaction strength $U_{d}$ to enter a nematic phase leads to unphysically large values.
For the nematic-spin-nematic phase, the respective interactions need to be even larger as can be seen in Fig.~\ref{fig:Upc}.
Finally, for the loop-current phase, we deduce values for the critical interaction strengths from Fig.~\ref{fig:varma}, that are much higher than realistically expected.
However, the aim of our analysis is not to answer whether this phases exist in the cuprates - a mean-field analysis would certainly not be the appropriate approach for such a task - but to analyze this different IUC orderings within the same framework.

We can also draw some conclusions about the competition or coexistence of the IUC-ordered phases from our calculation. Even though symmetry allows the two orders discussed in Sec.~\ref{sec:static}, the nematic and the nematic-spin-nematic, to coexist independently, they are promoted by different interactions, $V_{pp}$ and $U_p$, each hurting the respective other phase. In addition, both depend in a mean-field picture on the presence of a van Hove singularity. Even if the interactions were tuned in a way as to allow for both instabilities, having one kind of order removes already the high density of states from the Fermi level, thus preventing the system from entering the other phase. In contrast, the loop-current phase is promoted by the same interaction as the nematic phase, $V_{pp}$, and does not depend on a high density of states at the Fermi level. A deformation of the Fermi surface has thus no direct influence on this instability. Being of different symmetry, a nematic and a loop-current phase can therefore coexist in general independently.

The mean-field analysis and our exploration of the rich phase space of the three-band
Emery model in this paper can serve as a stepping stone
towards more sophisticated calculations of IUC orders and their
interdependence. For instance, extension of the calculations
in~\cite{halboth:2000, okamoto:2010} to the case of three bands might
provide further valuable insight. In particular, investigation of the
interplay between these IUC orders and superconductivity in a
genuinely strong coupling approach will be of great interest. As
superconductivity cannot be accessed within mean-field theory with
purely repulsive interactions, we left out this important issue
altogether~\footnote{Yamase and Metzner examined the competition
between nematic and superconductivity within a mean-field approach in the presence of attractive interactions~\cite{yamase:2007c}}. 

The possible coexistence of nematic and loop-current phases we find in
this work is interesting in light of experimental observations of
both IUC nematic order\cite{hinkov:2008, daou:2010} and  IUC staggered
magnetism in underdoped YBCO\cite{fauque:2006}. On the other hand, in Hg-compounds only
IUC staggered
magnetism has been observed\cite{li:2008}.
In order to test whether coexistence of both orders is a generic
feature, we propose a measurement of anisotropy
in Nernst effect in the presence of a symmetry breaking field on
Hg-compounds. For example, an in-plane magnetic field could align possibly existing nematic
domains in Hg-compounds which is tetragonal otherwise, much in the same
manner as in Sr$_3$Ru$_2$O$_7$\cite{borzi:2007}.

\appendix
\section{$t_{pp}=0$ analysis}
\label{app:tpp0}
In this appendix, the case $t_{pp}=0$ is investigated for $\eta\rightarrow 0$ for which we write the Hamiltonian~\eqref{eq:h0k} as 
\begin{equation}
  \HH_{\kk s} = \HH^{({\rm iso})}_{\kk s} + \eta \delta \HH_{\kk s},
  \label{eq:applinH}
\end{equation}
where
\begin{equation}
  \HH^{({\rm iso})}_{\kk s} = \left(\begin{array}{ccc} \xi^{({\rm iso})}_p & 0 & \gamma_{1}(k_x) \\ 0 & \xi^{({\rm iso})}_p & \gamma_{1}(k_y)\\ \gamma_{1}(k_y)& \gamma_{1}(k_y)&\xi^{({\rm iso})}_{d} \end{array}\right)
  \label{eq:sH0}
\end{equation}
and 
\begin{equation}
  \delta\HH_{\kk s} = \left(\begin{array}{ccc} -\frac{\tilde{V}_{pp}}{4} & 0 & 0 \\ 0 & \frac{\tilde{V}_{pp}}{4} & 0 \\ 0 & 0 &0 \end{array}\right)
  \label{eq:sdH}
\end{equation}
with 
\begin{eqnarray}
  \xi^{({\rm iso})}_p &=& \Delta + \tilde{U}_{p} \frac{n^{p}}{4} - \mu,\\
  \xi^{({\rm iso})}_{d} &=& \tilde{U}_{d} \frac{(n-n^p)}{2} - \mu.
  \label{eqn:diagtpp0}
\end{eqnarray}
The unperturbed Hamiltonian \eqref{eq:sH0} can straight-forwardly be diagonalized yielding the eigenenergies $\xi_{2 \kk s}^{({\rm iso})} = \xi^{({\rm iso})}_{p}$ and 
\begin{multline}
  \xi_{3/1 \kk s}^{({\rm iso})} =  \frac12(\xi^{({\rm iso})}_{p}\! + \xi^{({\rm iso})}_{d})\\
  \pm \sqrt{\frac14(\xi^{({\rm iso})}_{p}\! - \xi^{({\rm iso})}_{d})^2 + \gamma_{1}^2(k_x) + \gamma_{1}^{2}(k_y)}
  \label{eq:e0}
\end{multline}
with the corresponding states $|v_{\alpha}\rangle$ given by the eigenvectors
\begin{equation}
  \vec{v}_{1} = \left(\begin{array}{c}-\tilde{\gamma}_{1x}v_{\kk}\\-\tilde{\gamma}_{1y}v_{\kk}\\u_{\kk} \end{array}\right)\!, \;\vec{v}_{2} = \left(\begin{array}{c}-\tilde{\gamma}_{1y}\\\tilde{\gamma}_{1x}\\0\end{array}\right)\!, \; \vec{v}_{3} = \left(\begin{array}{c}\tilde{\gamma}_{1x}u_{\kk}\\\tilde{\gamma}_{1y}u_{\kk}\\v_{\kk}\end{array}\right). 
  \label{eq:vecs}
\end{equation}
In these equations, we introduced $u_{\kk}=\cos\frac{\omega_{\kk}}2$, $v_{\kk}=\sin\frac{\omega_{\kk}}2$ with
\begin{equation}
  \omega_{\kk} = \arctan\left(\frac{2\sqrt{\gamma_{1}^{2}(k_x) + \gamma_{1}^{2}(k_y)}}{\xi^{({\rm iso})}_{p} - \xi^{({\rm iso})}_{d}}\right)
  \label{eq:appomega}
\end{equation}
and 
\begin{equation}
  \tilde{\gamma}_{1x(y)} = \frac{\gamma_{1}(k_{x(y)})}{\sqrt{\gamma_{1}^2(k_x) + \gamma_{1}^{2}(k_y)}}.
  \label{eq:appgammatilde}
\end{equation}
For $\eta \rightarrow 0$, we can thus express the eigenenergies in powers of $\eta$ in a text-book perturbation-theory expansion,
\begin{equation}
  \xi_{\alpha \kk s} = \xi_{\alpha \kk s}^{({\rm iso})} + \eta \xi_{\alpha \kk s}^{(1)} + \eta^2 \xi_{\alpha \kk s}^{(2)} + O(\eta^3),
  \label{eq:expansion}
\end{equation}
with
\begin{equation}
  \xi_{\alpha \kk s}^{(1)} = \langle v_{\alpha}|\delta\HH|v_{\alpha}\rangle
  \label{eq:xi1}
\end{equation}
and 
\begin{equation}
  \xi_{\alpha\kk s}^{(2)} = \sum_{\beta\neq\alpha}\frac{|\langle v_{\alpha}|\delta\HH|v_{\beta}\rangle|^2}{\xi_{\alpha \kk s}^{({\rm iso})} - \xi_{\beta\kk s}^{({\rm iso})}}.
  \label{eq:xi2}
\end{equation}
The derivatives appearing in the self-consistency equations~\eqref{eq:selfconn}, \eqref{eq:selfconeta} and \eqref{eq:omegaetaeta} can thus all be expressed analytically through Eq.~\eqref{eq:expansion} yielding
\begin{eqnarray}
  \left.\frac{\partial \xi_{\alpha \kk s}}{\partial n^p}\right|_{\eta=0} &=& \frac{\partial \xi_{\alpha \kk s}^{({\rm iso})}}{\partial n^p}\label{eq:xinp},\\
  \left.\frac{\partial \xi_{\alpha \kk s}}{\partial \eta}\right|_{\eta=0} &=& \xi_{\alpha \kk s}^{(1)}\label{eq:xieta},\\
  \left.\frac{\partial^2 \xi_{\alpha \kk s}}{\partial \eta^2}\right|_{\eta=0} &=& 2 \xi_{\alpha \kk s}^{(2)}\label{eq:xietaeta}.
\end{eqnarray}
Evaluating the derivatives for the oxygen hole density in Eq.~\eqref{eq:selfconn}, we find using
\eqref{eq:xinp}
\begin{eqnarray}
  \left.\frac{\partial \xi_{1 \kk s}}{\partial n^p}\right|_{\eta=0} &=& (\frac{\tilde{U}_{p}}{4} + \frac{\tilde{U}_{d}}{2}) v_{\kk} - \frac{\tilde{U}_{d}}{2},\\
  \left.\frac{\partial \xi_{2 \kk s}}{\partial n^p}\right|_{\eta=0} &=& \frac{\tilde{U}_{p}}{4} = \frac{\tilde{U}_{p}}{4} + \frac{\tilde{U}_{d}}{2} - \frac{\tilde{U}_{d}}{2},\\
  \left.\frac{\partial \xi_{3 \kk s}}{\partial n^p}\right|_{\eta=0} &=& (\frac{\tilde{U}_{p}}{4} + \frac{\tilde{U}_{d}}{2}) u_{\kk} - \frac{\tilde{U}_{d}}{2}.
  \label{eq:explicitxinp}
\end{eqnarray}
Using these derivatives, the self-consistency equation for the oxygen occupation number simplifies to
\begin{multline}
  n^p = \frac{1}{N}\sum_{\kk, s}\left\{v_{\kk}^2n_{\rm F}(\xi_{1\kk s}^{({\rm iso})})\right.\\
  \left.+ n_{\rm F}(\xi_{2\kk s}^{({\rm iso})}) + u_{\kk}^2 n_{\rm F}(\xi_{3\kk s}^{({\rm iso})})\right\}
  \label{eq:appnpeta0}
\end{multline}

\section*{Acknowledgements}
We thank Steve Kivelson, Michael Lawler, Kyungmin Lee, Richard
Scalettar, Eduardo Fradkin, Walter Metzner, Hiroyuki Yamase, Thomas
Maier, Shiquan Su, Carsten Honerkamp, and Andre-Marie Tremblay for useful discussions.
We acknowledge support from NSF Grant DMR-0520404 to the Cornell Center for Materials Research and from NSF Grant DMR-0955822.

\end{document}